\documentclass{aa}

\usepackage{graphicx}
\usepackage{natbib}
\bibpunct{(}{)}{;}{a}{}{,}

\begin{document}

\hyphenation{Fe-bru-ary Gra-na-da mo-le-cu-le mo-le-cu-les}
%
%     \title{Estimation and reduction of the uncertainties in chemical models: Application to dark cloud chemistry}
	\title{The effect of uncertainties on chemical models of dark clouds}

     \subtitle{}

     \author{V. Wakelam\inst{1} , E. Herbst\inst{1,2}, F. Selsis\inst{3}}
     \institute{ Department of Physics, The Ohio State University, Columbus, OH 43210, USA \and Departments of Astronomy and Chemistry, The Ohio State University, Columbus, OH 43210, USA\and Centre de Recherche Astronomique de Lyon, Ecole Normale Sup\'erieure, 46 All\'ee d'Italie, F-69364 Lyon cedex 7, France       }
     \offprints{wakelam@mps.ohio-state.edu}
     \date{Received xxx / Accepted xxx }
     
     \abstract{ The gas-phase chemistry of dark clouds has been studied with a treatment of uncertainties caused both by errors in individual rate coefficients and uncertainties in physical conditions.  Moreover, a sensitivity analysis has been employed to attempt to determine which reactions are most important in the chemistry of individual species.  The degree of overlap between calculated errors in abundances and estimated observational errors has been used as an initial criterion for the goodness of the model and the determination of a best `chemical' age of the source.  For the well-studied sources L134N and TMC-1CP, best agreement is achieved at so-called ``early times'' of 
 $\approx$ 10$^{5}$ yr , in agreement with previous calculations but here put on a firmer statistical foundation. A more detailed criterion for agreement, which takes into account the degree of disagreement, is also proposed.  Poorly understood but critical classes of reactions are delineated, especially reactions between ions and polar neutrals.  Such reactions will have to be understood better before the chemistry can be made more secure. Nevertheless, the level of agreement is low enough to indicate that a static picture of physical conditions without consideration of interactions with grain surfaces  is inappropriate for a complete understanding of the chemistry.
  }

     \titlerunning{Chemical modeling uncertainties}
     \authorrunning{Wakelam et al.}

     \maketitle

\section{Introduction}

The chemistry of dark clouds, now known as quiescent cores,  has been studied for over thirty years \citep{1973ApJ...185..505H,1973ApJ...183L..17W}.  Indeed, a significant fraction of the more than 130 gas-phase species detected via high-resolution spectroscopy in the interstellar and circumstellar media has been seen in such sources, which include the well-studied clouds TMC1-CP and L134N \citep{2004MNRAS.350..323S,1998ApJ...501..207T}.  In an attempt to reproduce the abundances of these species, a large number of gas-phase chemical models have been developed that compute the variation of the gas-phase concentrations as a function of time.  For many years, most of these models were based on the pseudo-time-dependent approximation \citep{1980ApJS...43....1P,1984ApJS...56..231L,1984ApJ...277..581L,1985MNRAS.217..507M}, in which the chemistry evolves under fixed and homogeneous conditions from some initial abundances,  consisting of  totally atomic material except for a high abundance of H$_{2}$. In the last decade, more complex models, including surface chemistry \citep{1992ApJS...82..167H,2000MNRAS.319..837R}, heterogeneity of the source \citep{2001A&A...369..605H}, and assorted effects of dynamics and turbulence \citep{2000ApJ...535..256M}, have been reported.   

In all of these models, the connections among gas-phase species are described by up to thousands of chemical reactions, with rates quantified by rate coefficients, most of which are poorly determined by experimental or theoretical means. Nevertheless, even the older pseudo-time-dependent models were at least partially successful in reproducing the exotic nature of the chemistry (molecular ions, radicals, and metastable isomers), the unsaturated nature of the more complex products (e.g., cyanopolyynes), and the strong deuterium fractionation.  They have been less successful, however, in detailed comparisons with observations of fractional abundances and column densities for the large number of species observed in individual sources.  In the last published comparison between a pseudo-time-dependent theory and observation for the well-studied source TMC1-CP \citep{2004MNRAS.350..323S}, it was found that at the time of best agreement ($1 \times 10^{5}$ yr), only about 2/3 of the observed species have calculated fractional abundances within an order-of-magnitude of the observed values.  This result, obtained with the relatively new osu.2003 network and so-called ``low-metal'' elemental abundances, is actually worse than obtained with a previous network from the Ohio group, known as the ``new standard model'' (nsm), where nearly 4/5 of the molecular abundances were reproduced to within one order of magnitude \citep{1998ApJ...501..207T}.   The order-of-magnitude criterion used in both studies is a subjective one, however,  and no comparison involving TMC1-CP has ever been done by taking into account the rate coefficient uncertainties in the computation of the theoretical abundances.  

Without any quantification of the random model error, it is difficult to conclude if observed abundances can be reproduced or not by the model.
Indeed, including the estimated uncertainties in rate coefficients is the only way to define those species with abundances not reproduced by chemical networks.  With a rigorous quantitative approach to uncertainty, moreover, one can begin to comprehend the deficiencies of gas-phase pseudo-time-dependent calculations, and determine where improvement is most necessary.  If, for example,  hydrogen-rich species, which may be synthesized on grain surfaces and desorbed into the gas via non-thermal means, are the only species with abundances to be poorly reproduced, the most important correction would involve the inclusion of surface processes.  If, on the other hand, smaller species such as O$_{2}$ and H$_{2}$O, with simple chemistry, cannot be reproduced within the estimated errors, then it is at least conceivable that a far more complex physical scenario is needed in which the formation and dynamics of the quiescent core play roles in the chemistry (Bergin \& Melnick 2005).  Whether or not to give up on the pseudo-time-dependent approximation then depends critically on our estimates of uncertainties, because there is no point in violating Occam's Razor if a more simple theory is as adequate as a more complex one.  

This paper is the second application of a method developed in \citet{Wakelam2005} to quantify the random errors of chemical models. In \citet{Wakelam2005}, we presented the method in great detail  and showed some consequences for hot core chemistry (T$> 100$~K).  Hot cores occur during an evolutionary stage of star formation, and can have complex structures not well understood  \citep{2004A&A...414..409N}. Dark clouds, which have not yet started to collapse in an appreciable manner,  may, on the contrary,  have relatively simple temperature and density structures. These objects are then better laboratories to test the gas-phase chemical models usually used.  In this paper, we apply the method to estimate the statistical errors as a function of time for dark cloud chemistry, which is occurring at  an H$_2$ density of $\sim 10^4$~cm$^{-3}$ and a temperature $< 20$~K.  A previous analysis of uncertainties for mainly steady-state conditions was performed for dark clouds by \citet{2004AstL...30..566V}, although their error was defined in a different manner.  

The  remainder of our paper is organized as follows.  In Section 2, we briefly describe the chemical network and approach to uncertainties used for this study. Some general results are presented in Section 3, where we also (i) consider differences between the two major networks, and (ii) make a comparison between our uncertainty calculations and those of \citet{2004AstL...30..566V}.  Section 4 contains a comparison between our results and observed abundances towards the two dark clouds L134N and TMC1-CP, while  Section 5 discusses the specific cases of O$_2$ and H$_2$O, and the cloud ages. We conclude the paper in Section 7.

\section{Chemical network and uncertainty method}\label{model}

\begin{figure}
\begin{center}
\includegraphics[angle=90,width=1\linewidth]{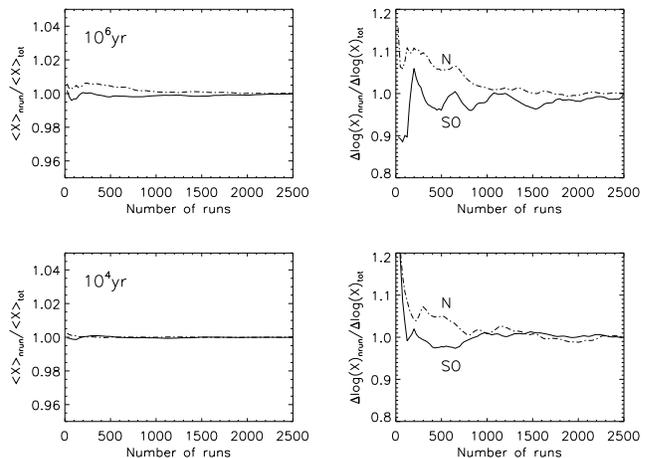}
\caption{Mean abundance ($<X>$) (left panels) and error $\Delta \log X$ (right panels) for N and SO plotted as a function of the number of runs.  Both quantities are normalized  by the value obtained for the maximum number of runs (2500). The results are given for Model 2 at  two times: $10^4$~yr on the lower panels and $10^6$~yr on the upper panels.  }
\label{ntir}
\end{center}
\end{figure}

\begin{table}
\caption{Physical parameters for the models.\label{diff_mod}}
\begin{tabular}{ll}
\hline
\hline
\multicolumn{2}{c}{Physical conditions} \\
\hline
Model 1 & 10~K, $10^4$~cm$^{-3a}$ \\
Model 2 & 5-15~K, $(0.5-1.5)\times 10^4$~cm$^{-3a}$  \\
Model 3 & 5-15~K, $(0.5-1.5)\times 10^4$~cm$^{-3a}$, C/O=1.2 \\
\hline
\end{tabular}
\begin{list}{}{}
\item $^a$ H$_2$ density. 
\end{list}
\end{table}%

\begin{table}
\caption{Initial elemental abundances with respect to H$_2$.\label{init}}
\begin{tabular}{llll}
\hline
\hline
\multicolumn{4}{c}{Initial abundances}  \\
\hline
He & $2.8\times 10^{-1}$ & Fe$^+$ & $6.0\times 10^{-9}$\\
N & $4.28\times 10^{-5}$ & Na$^+$ & $4.0\times 10^{-9}$\\
O & $3.52\times 10^{-4}$ & Mg$^+$ & $1.4\times 10^{-8}$ \\
C$^+$ & $1.46\times 10^{-4}$ & P$^+$ & $6.0\times 10^{-9}$\\
S$^+$ & $1.6\times 10^{-7}$ & Cl$^+$ & $8.0\times 10^{-9}$\\
Si$^+$ & $1.6\times 10^{-8}$ & & \\
\hline
\end{tabular}
\end{table}

We used a time-independent physical model with the gas-phase chemical network osu.2003\footnote{ http://www.physics.ohio-state.edu/$\sim$eric/research.html} reported by \citet{2004MNRAS.350..323S}. This network contains standard gas-phase reactions (ion-neutral, neutral-neutral, dissociative recombination etc), with the addition of a significant number of rapid radical-neutral processes. Except for H$_{2}$ production, the grains are only important as sites of negative charge that recombine with positive ions.  We considered the uncertainties in all the classes of reactions except for ion recombination with negative grains. The uncertainties in the rate coefficients have recently been added to the osu.2003 network, based on those listed in the RATE99 network \citep{2000A&AS..146..157L}\footnote{ http://www.udfa.net}. For the uncertainties not estimated  at 10~K (according to the temperature range given in RATE99), we increased the uncertainty to a factor  of 2. 

The Monte Carlo method used here to include the rate-coefficient uncertainties is described in detail by \citet{Wakelam2005}. Briefly, it consists of generating $N$ new sets of rate coefficients by replacing each coefficient $k_i$ by a random value consistent with its uncertainty factor $F_i$.   We assume a normal distribution of $\log k_i$ with a standard deviation $\sigma_i = \log F_i$.  We run the model for each set $j$, which produces, for each species, $N$  values of the abundance $X_j (t)$ at a time $t$. The mean value of $\log X(t)$ gives us the ``recommended" value while the dispersion of $\log X_{j}(t)$ around  $<\log X(t)>$ determines the error due to kinetic data uncertainties. The error in the abundance $ \Delta \log X = \frac{1}{2}( \log X_{\mathrm{max}} - \log X_{\mathrm{min}})$ is defined at a time $t$ as the smallest interval $[\log X_{\mathrm{min}},\log X_{\mathrm{max}}]$ that contains 95.4\% of $\log X_j$ values, which is equivalent to $2\sigma$ for symmetric Gaussian distributions.  For example, if $ \Delta \log X = 1.0$ and the calculated distribution is symmetric, then  the $2\sigma$ values of $X_{\mathrm{max}}$ and $X_{\mathrm{min}}$ are one order of magnitude greater and less than $X$, respectively.  Similarly, if $ \Delta \log X = 0.5$, the $2\sigma$ values are greater and less than $X$ by a factor of 3.3, and if 
$ \Delta \log X = 0.3$, they are greater and less than $X$ by a factor of 2.0.

We initially ran two models with different physical conditions. In the first model (Model 1), the temperature and the H$_2$ density are held constant at values of  10~K and $10^4$~cm$^{-3}$, respectively;  only the rate coefficients vary.  For the second model (Model 2), we randomly chose the temperature and the density within a possible range of values. This uncertainty was adopted for two reasons: (i)  the cloud temperature and density are usually derived from observations (using approximate line excitation models) and an error in these values can usually be determined, and (ii) the modeled clouds are more likely to be inhomogeneous  sources with contributions over the chosen ranges of temperature and density. We investigated the sensitivity of the results if an uncertainty of $\pm$50\% is considered for T and n(H$_2$) around the typical values given for Model 1 (see Table~\ref{diff_mod}). Because temperature and density are physical parameters not characterized by a Gaussian distribution, we used a flat distribution instead for the rate coefficients.  
For both models, we used a cosmic-ray ionization rate of $1.3\times 10^{-17}$~s$^{-1}$, a visual extinction of 10 and the low-metal elemental abundances, which are listed in Table~\ref{init}. A third model, with carbon-rich elemental abundances, is discussed later in the text.

For Model 1 and Model 2, we performed 2000 and 2500 different runs, respectively. In order to demonstrate convergence,  we plot as examples in Fig.~\ref{ntir} the mean abundance $<X>$ and its error  $\Delta \log(X)$ for the species N and SO as functions of the number of runs at two different times with Model 2.  The plotted parameters are normalized by the values obtained for 2500 runs.   For all the species, we noticed that the number of runs chosen is more than adequate. 

\section{Calculated uncertainties in abundances}

\subsection{General considerations}

\begin{figure}
\begin{center}
\includegraphics[width=1\linewidth]{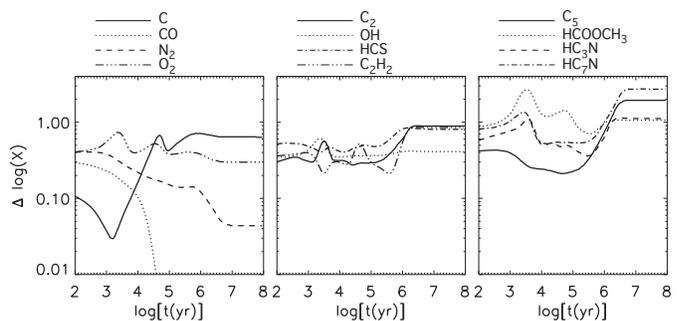}
\caption{Error $\Delta \log X$ as a function of time for some species with Model 1.}
\label{sigma_time}
\end{center}
\end{figure}

\begin{figure}
\begin{center}
\includegraphics[width=0.8\linewidth]{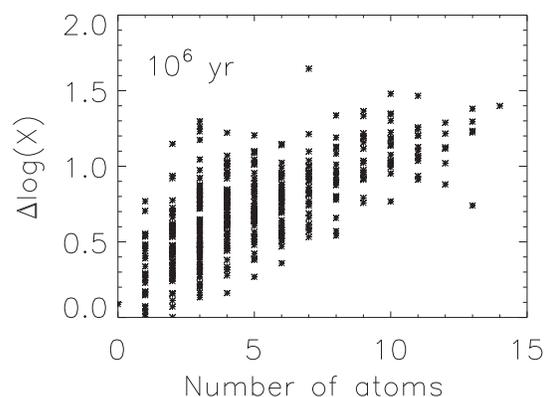}
\caption{Error, $\Delta \log X$, as a function of the number of atoms per molecule at $10^6$~yr for Model 1.}
\label{err_atome}
\end{center}
\end{figure}

Figure~\ref{sigma_time} presents the error $\Delta \log X$ for some species as a function of time and for the physical conditions of Model 1 (see Table~\ref{diff_mod}). As already noticed for hot core chemistry, the errors in the abundances  increase with time for most of the molecules except the species that contain a dominant portion of an element, such as CO for  carbon and N$_2$ for  nitrogen.  For example, the errors for  HCN and HC$_3$N are 0.2 and 0.5 at $10^5$~yr and increase to 0.3 and 1.1, respectively, at $10^7$~yr. The error also increases with the complexity of the molecule, an effect previously noted by \citet{2004AstL...30..566V}.  This effect is shown in Fig.~\ref{err_atome} for the physical parameters of Model 1 at a time of 10$^{6}$ yr.  For molecules with 10 atoms, the figure shows that an error of 1.0, corresponding to a factor of 10, is in the middle of the range, while for molecules with five atoms, the median error is $\approx$ 0.7, corresponding to a factor of five.

\subsection{Variation of the temperature and density}\label{var_Tn}

\begin{figure}
\begin{center}
\includegraphics[width=1.\linewidth]{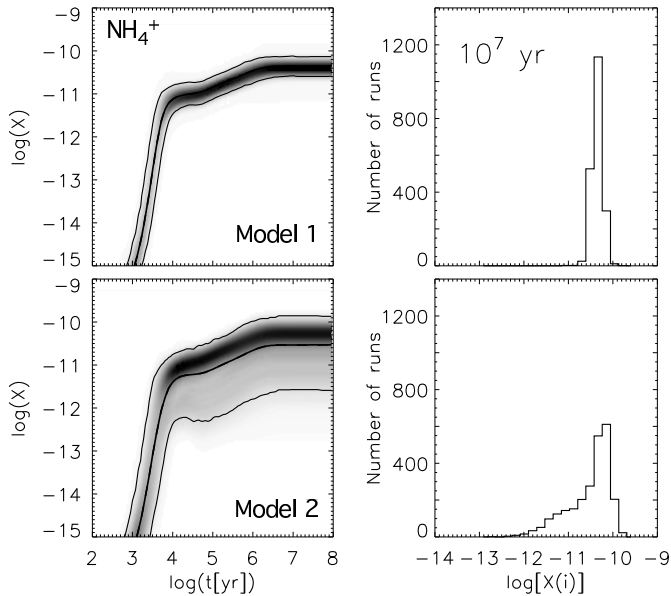}
\caption{Density of probability of the NH$_4^+$ abundance. The right plots represent the histograms of the abundance distributions at $10^7$~yr. The top row is for Model 1 and the bottom for Model 2.  }
\label{density_dist_NH4p}
\end{center}
\end{figure}

\begin{figure}
\begin{center}
\includegraphics[width=0.8\linewidth]{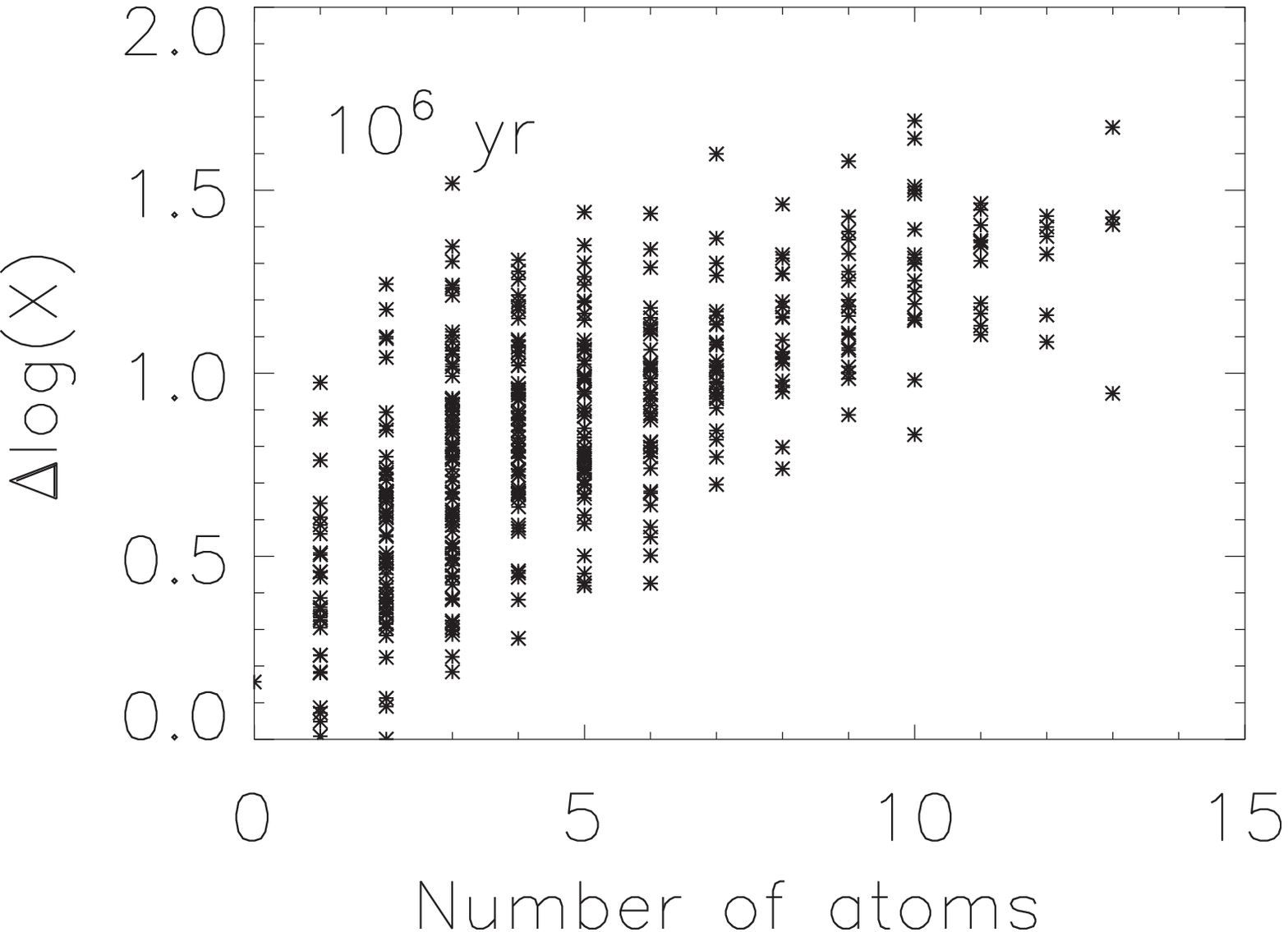}
\caption{Error, $\Delta \log X$, as a function of the number of atoms per molecule at $10^6$~yr for Model 2.}
\label{err_mod2_atome}
\end{center}
\end{figure}

\begin{table}
\caption{Examples of errors computed using Model 1 ($\Delta \log X_1$) and Model 2 ($\Delta \log X_2$) at $10^5$~yr.  \label{ex_error12}}
\begin{tabular}{lll}
\hline
\hline
Species & $\Delta \log X_1$ & $\Delta \log X_2$\\
\hline
N$^+$ & 0.36 & 1.0 \\
HNO & 0.31 & 0.91 \\
NH$_3$ & 0.27 & 0.88 \\
NH$_2$CN & 0.37 & 0.87 \\
SO & 0.64 & 0.79 \\
H$_2$O & 0.40 & 0.42 \\
HC$_3$N & 0.45 & 0.56 \\
C$_9$N & 0.43 & 0.53 \\
\hline
\end{tabular}
\end{table}%

The variations of the gas temperature and density  in Model 2 cause two major effects.  First, they modify the distribution of the abundances at a given time from a Gaussian shape to an asymmetrical one, as can be seen by the example of NH$_4^+$ in Fig.~\ref{density_dist_NH4p}.  The effect is not surprising since the rate equations are not linear with respect to density or temperature.  Indeed, the dependence on temperature can be exponential for processes with a small barrier.
%Note that contrary to \citet{2004AstL...30..566V} we do not find an asymmetrical profile if we keep the temperature and density constant because the shape of their histograms is due to the uniform distribution of $K$ uncertainties that they used. 
The second effect is an increase in the error, which can be significant for some of the species,  as shown in Fig.~\ref{err_mod2_atome}, which is to be compared with Fig.~\ref{err_atome}. The nitrogen-bearing species are particularly sensitive to the variation of temperature and density, as can be seen by the examples in Table \ref{ex_error12}.  What causes this extreme sensitivity? It is likely that the variation in temperature is the cause
  because the nitrogen chemistry starts with the reaction N$^+$ + H$_2$ $\rightarrow$ NH$^+$ + H, which is assumed to be endothermic by 85~K in both the osu.2003 and RATE99 networks.   The situation is more complex than can be contained in networks because it is likely that an estimated non-thermal fraction of H$_{2}$ in its $J$=1 $ortho$ state of 10$^{-3}$ drives the reaction and essentially converts all the atomic nitrogen ion into NH$^{+}$ \citep{1991A&A...242..235L}.  Nevertheless, with the adopted rate coefficient, the reaction to form NH$^{+}$ is so inefficient at temperatures under 10 K that it loses importance. Thus, temperatures lower than 10~K lead to much lower abundances for some N-bearing species.   The result is a skewing of the abundance distributions to lower values, as can be seen for the case of NH$_{4}^{+}$ in Fig. \ref{density_dist_NH4p}.  The effect of the temperature and density variations is less important for complex molecules because the variations caused by the physical parameters are diluted by the large uncertainties due to the uncertainties in rate coefficients. For example, the uncertainties in the cyanopolyne abundances do not show a strong dependence on the physical parameters: the increase of $\Delta \log X$ with Model 2 is less than a factor of 2 for HCN and this number decreases with the increasing complexity of the cyanopolyne.  
 
\subsection{Comparison between osu.2003 and RATE99 databases}

\begin{figure}
\begin{center}
\includegraphics[angle=0,width=0.8\linewidth]{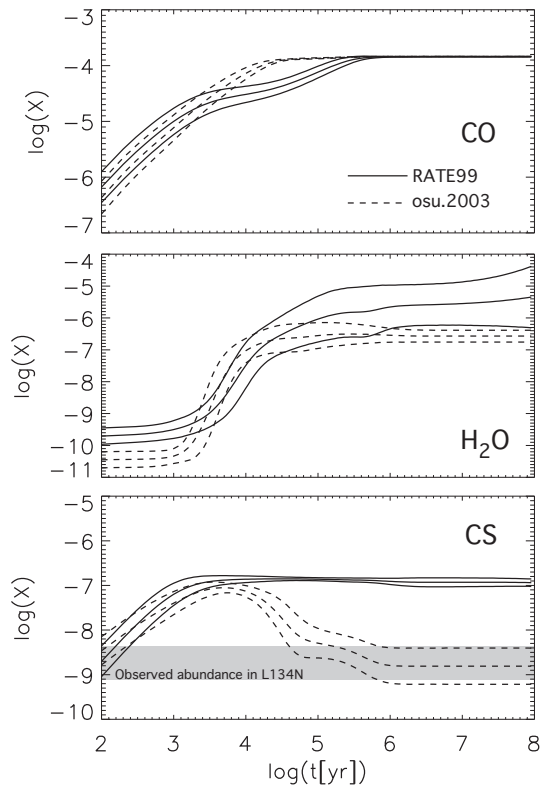}
\caption{CO, H$_2$O and CS abundances as a function of time computed using the osu.2003 (dashed lines) and the RATE99 (solid lines) databases. The three curves for each molecule and database refer to the average value and the $2\sigma$ errors.}
\label{osu_umist}
\end{center}
\end{figure}

The statistical uncertainties in the rate coefficients are not the only sources of error for gas-phase models.  Most of the reactions have not been studied in the laboratory or via detailed theoretical considerations, so that approximate values for the rate coefficients must be used.  Even studied reactions, if studied by more than one group, show large discrepancies in rates.  Some of the problems in choosing proper rate coefficients can be seen in a comparison of the osu.2003 and RATE99 databases, which contain many reactions with different choices of rate coefficients. These differences arise from several specific sources:
\begin{itemize}
\item[a)] Different estimates for poorly understood reactions such as radical-neutral reactions and  both neutral-neutral and ion-molecule radiative associations,
\item[b)] Different choices of experimental values from uncritical compilations such as the NIST Chemical Kinetics Database (http://kinetics.nist.gov/index.php) or from competing measurements,
\item[c)] Different approximations regarding the temperature dependence of ion-polar neutral 
reactions.
\end{itemize}
The problem with radical-neutral reactions has been studied by \citet{2004MNRAS.350..323S}: the osu.2003 network includes estimates of rapid rates for a variety of such reactions that are greater than used in RATE99 as well as the previous network from the Ohio group.  The last discrepancy derives from the fact that although experiments show that virtually all of the small number of ion-polar neutral reactions studied possess an inverse temperature dependence \citep{1992IAUS..150....7R,1992Rowe}  
this dependence is not easily expressible in terms of the simple rate expression used in the networks.  Based on the work of \citet{1986ApJ...310..378H}, the osu.2003 network uses an approximation derived from  the locked dipole approximation for linear neutrals and classical scaling approach for non-linear neutrals \citep{1982Su}. Both of these approximations lead to a temperature dependence of T$^{-1/2}$, while the RATE99 network assumes the rate coefficients to have no temperature dependence.  More detailed studies show that the inverse dependence of the rate coefficient on temperature may well be in between these two limiting cases \citep{2005ApJ...628..260N}. More work is clearly needed on these systems.

At 10~K, 60\% of the reactions present in both databases show a difference in rate coefficient lower than the considered uncertainty, a condition which can be expressed as
\begin{equation}
 \frac{{ k}_{i1}/k_{i2}}{\sqrt{2} F_i} \le 1,
 \end{equation}
 where ${ k}_1$ and $k_2$ are the rate coefficients from the different networks with $ k_1 > k_2$ and $F$ is the uncertainty factor.  In other words, almost half of the common reactions show a difference in rate coefficient not covered by the uncertainties in the rate coefficients, with 2\% of them having an ``extra" difference greater than three orders of magnitude ($\frac{{k}_{i1}/k_{i2}}{\sqrt{2} F_i} > 1000$). These differences may not have strong consequences on the modeling results, however, if the reactions concerned are not important or if formation and depletion reactions in one model are both larger to a similar extent than in the other.  So, one must ask whether or not the differences between the abundances computed with the two databases are larger than the errors due to rate coefficient uncertainties.

To answer this question, we ran Model 1 with the RATE99 network so as to compare with the osu.2003 results. We obtained distributions of results for abundances that do not always overlap at the  $2\sigma$ level.  In fact,  eighty-four percent of the 373 common species show a disagreement at some times between the abundances computed with the two databases. No neutral species and only 62 ions, such as S$^+$, C$_2^+$, CN$^+$ and C$_6^+$, overlap at all times. Note that the differences are due to difference in the values of the rate coefficients and not in the differing reaction lists for the two networks, since we repeated the comparison with a list of reactions common in both databases and obtained similar results. Fig.~\ref{osu_umist} shows three examples: the CO, H$_2$O and CS molecules. The error in the abundance of CO at later times is very low because this molecule is the main reservoir of atomic carbon in the gas phase and both sets of reactions give the same results. At early times, between $6\times 10^3$ and $2\times 10^5$~yr, the abundance computed with osu.2003 is higher than the one computed with RATE99, reaching a ratio of 4 at $3\times 10^4$~yr.  The envelopes defined by the error in the rate coefficients do not overlap. For H$_2$O, the early and late abundances diverge by one order of magnitude and the envelopes do not quite overlap, whereas CS is produced to a much greater extent for the RATE99 case: the difference of up to 2 orders of magnitude at steady state far exceeds the envelopes of error. Indeed, with RATE99, CS is found to be the reservoir of S: the difference between the 2 networks is thus quite profound and not only quantitative.

 We attempted to identify the reason for the discrepancy involving CS using the sensitivity method described in \citet{Wakelam2005}. At $10^5$~yr, of the 20 most important reactions in the calculation of CS with the osu database, 10 are different in RATE99 and for those 10, 7 are ion- polar neutral reactions. If we replace these 10 rate coefficients by their values in RATE99, the CS result approaches the RATE99 abundance to within a factor of $\approx$ 3. Thus, one would argue that the sensitivity approach works well in deducing both the ``important'' reactions in the osu calculation and the reason for the difference with the results from the RATE99 calculations.  But, one must be cautious because different results are obtained if one starts with RATE99: here,
because of their smaller rate coefficients, ion-polar neutral reactions play a less significant role in the production and destruction of CS. Indeed, the reactions of importance for CS are quite different in the two networks. The introduction of the osu rates for a small number of  "important" reactions in the RATE99 network produces only a small change although the difference in the rate coefficients of these reactions is not covered by the uncertainty factor. The RATE99 abundance of CS is thus more stable against rate coefficient modifications, which shows that a large number of reactions would have to be modified in order to change the result. 

 For CO, which shows a factor of 3 difference at early times, we were not able to identify any main reactions responsible for this discrepancy. Once again, these small differences are then due to a large number of reactions with small variations of rate coefficients between the two databases. 

\subsection{Comparison of uncertainty calculations}

\citet{2004AstL...30..566V} did a similar study on dark cloud chemistry to ours using a previous version of the RATE99 database with a flat distribution for the rate errors \citep[see ][for a discussion concerning this point]{Wakelam2005} and focused their results on the steady-state abundances. Although they defined the error in the abundance in a different way, we can compare their results with ours at $10^8$~yr. The authors used the full width at half maximum (FWHM) of the histograms of $\log X$, which is $\sim \frac{2.35}{2} \Delta \log X$, as defined in Sect.~\ref{model}, since most of the histograms have a Gaussian shape \citep[see the discussion about the definition of the error in][]{Wakelam2005}. 
 For most of the species, we found a lower uncertainty than did \citet{2004AstL...30..566V}. As an example, the C$_4$ molecule has a $\Delta \log X$ of 1.45 at $10^8$~yr in our study whereas \citeauthor{2004AstL...30..566V} found an FWHM of 2-3 for carbon clusters. 
The reason for this difference appears to be mainly due to the database used, because  we found that the errors are generally higher with RATE99 than with osu.2003, as can be seen in Fig.~\ref{osu_umist}  for  the case of  H$_2$O.  For the specific case of the molecule C$_{4}$, we calculate that $\Delta \log X$ is 2  at $10^8$~yr with RATE99, a value well above what we calculate with osu.2003.  The source of this difference in errors is unclear.

\section{Comparison with observations}\label{Comp_obs}

\begin{table}
\caption{Observed abundances towards L134N.     \label{ab_L134N}}
\begin{center}
\begin{tabular}{lcclcc}
\hline
\hline
Species & $N(i)/N({\rm H}_2)$ & Ref & Species & $N(i)/N({\rm H}_2)$ & Ref \\
\hline
CH   &  1(-8) & (1)   & CN   &  8.2(-10)    & (1)\\
CO  &   8.(-5) &  (1) &    CS  &   1.7(-9) & (2) \\    
NO  &   6.(-8)  & (1)  &     OH  &   7.5(-8)  & (1)  \\    
SO   &  3.1(-9)   & (2)  & C$_2$H  &  $\le 5.(-8)$   & (1) \\
C$_2$S  &  6.(-10)  & (1) &   H$_2$S  &  8.(-10) & (1) \\    
HCN  &  1.2(-8)  & (2) & HNC  &  4.7(-8)  & (2) \\  
OCS  &  2.(-9)  & (1) & SO$_2$  &  $\le 1.6(-9)$  & (2)  \\
C$_3$H  &  3.(-10)  & (1) &      C$_3$N  &  $\le 2.(-10)$  & (1) \\
C$_3$O  &  $\le 5.(-11)$  & (1) &  C$_3$S &   $\le 2.(-10)$  & (1) \\
H$_2$CO  & 2.(-8) & (1) &     H$_2$CS &  6.(-10)  & (1) \\    
NH$_3$  &  9.1(-8)  &  (1) &     CH$_2$CN & $\le 1.(-9)$ & (1) \\
C$_2$H$_2$O & $\le 7.(-10)$  & (1) & C$_3$H$_2$ &   2.(-9) & (1) \\    
C$_4$H  &  1.(-9) & (1) &  HCOOH &  3.e-10 & (1)  \\    
HC$_3$N  &  8.7(-10)  & (2) &  CH$_3$CN  & $\le 1.(-9)$  & (1)\\
CH$_3$OH  & 3.7(-9)  & (2) &  CH$_3$CHO  & 6.(-10)  &(1) \\   
C$_2$H$_3$CN  & $\le 1.(-10)$  & (1)  &  C$_3$H$_4$ &   $\le 1.2(-9)$  & (1) \\  
HC$_5$N &  1.(-10)  &  (1) &  HC$_7$N &  2.(-11)  &  (1) \\    
HCO$^+$ &   1.(-8)   & (2)  &  HCS$^+$  &  6.(-11)  & (1)  \\    
N$_2$H$^+$ &  6.8(-10)  & (2) & H$_2$CN$^+$ &  $\le 3.1(-9)$  & (1)  \\
H$_2$O  &  $\le 3.(-7)$   &  (3) &  O$_2$  &   $\le 1.7(-7)$  &  (4) \\
C$^0$ & $\ge 1.(-6)$ & (5) & & & \\
\hline
\end{tabular}
\end{center}
$^{a}$  $a(-b)$ refers to $a \times 10^{-b}$ \\
(1) \citet{1992IAUS..150..171O} \\
 (2) \citet{2000ApJ...542..870D}\\
 (3) \citet{2000ApJ...539L.101S}\\
 (4) \citet{2003A&A...402L..77P} \\
 (5) \citet{1996A&A...311..282S}\\
\end{table}

\begin{figure}
\begin{center}
\includegraphics[width=0.8\linewidth]{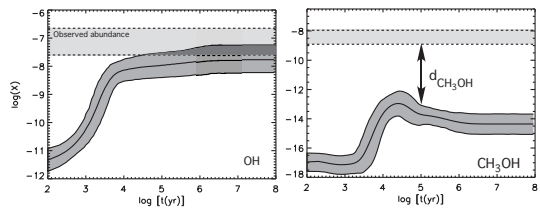}
\caption{Theoretical abundance of OH (left) and CH$_3$OH (right) as a function of time for Model 2. The dashed lines represent the observed values in L134N with an error of a factor of 3. The agreement between the observed and modeled abundance of OH is shown as a dark area whereas the disagreement for CH$_3$OH is quantified by a distance of disagreement d (see Sect.~\ref{cloud_estimate}).}
\label{Comp_L134N_OH}
\end{center}
\end{figure}

We have compared our theoretical results with some observations in two dark clouds: L134N (N position) and TMC-1 (CP peak). The  abundances observed in L134N are summarized in Table~\ref{ab_L134N} while those observed in TMC-1CP are listed in \citet[][Table 3]{2004MNRAS.350..323S}. The goal of this comparison is to take into account both the uncertainty in the observed values and in the chemical modeling. For the observed abundances, since all the abundances are not given with their corresponding errors, we assumed a standard uncertainty of  $\pm$ a factor of 3 for the following reasons. First, telescope and atmospheric calibrations are responsible for $\sim$ 30\% of uncertainty in the observed abundances. In addition, the abundances relative to H$_2$ depend on the H$_2$ column density, which is an additional source of uncertainty.  The H$_2$ column density is usually indirectly determined from the emission of other molecules via LTE, LVG, or Monte-Carlo models and usually different molecules give different estimates of H$_2$ densities. Indeed,  the emission of the molecules may come from different the layers of gas. In TMC-1CP for instance, it appears that C$^{18}$O may not come from the same volume of gas as the other molecules  and  the abundances compared to H$_2$ may be overestimated \citep{1997ApJ...486..862P}.  Another reason for the high uncertainty of the observed values is that the inventory of the observed abundances come from several studies, in which  several telescopes and approximations were used to compute the observed abundances, resulting in disagreements by at least a factor of three for some species.  As a specific example, the SO abundance towards L134N (N position) is $3.1\times 10^{-9}$ in \citet{2000ApJ...542..870D} whereas it is 6 times higher in \citet{1992IAUS..150..171O}. Even though we think that an uncertainty of $\pm$ a factor of 3 is reasonable, and consider it to be our standard observational uncertainty, we also use a larger uncertainty of $\pm$ a factor of 5 to consider its effect.

For the chemical model, we considered both the uncertainties in the rate coefficients and in the physical conditions, using Model 2 for both clouds.  To compare the observed and modeled values, we first define  agreement between the model and the observations to occur when the error bars of the calculated and observed values overlap.  A more detailed approach is discussed below in Section 
\ref{cloud_estimate}.  An example of overlapping error bars at certain times only can be seen in Fig.~\ref{Comp_L134N_OH} for the case of OH in L134N, where the overlap is shown as a darker region.  
%{\bf The maximum faction of species reproduced by the different models (Models 1 to 3) and using the two databases osu.2003 and rate99 are summarized in Table~\ref{sum_fact} together with the mean time of "best" reproduction. } 

\subsection{Results for L134N}

\begin{figure}
\begin{center}
\includegraphics[width=1\linewidth]{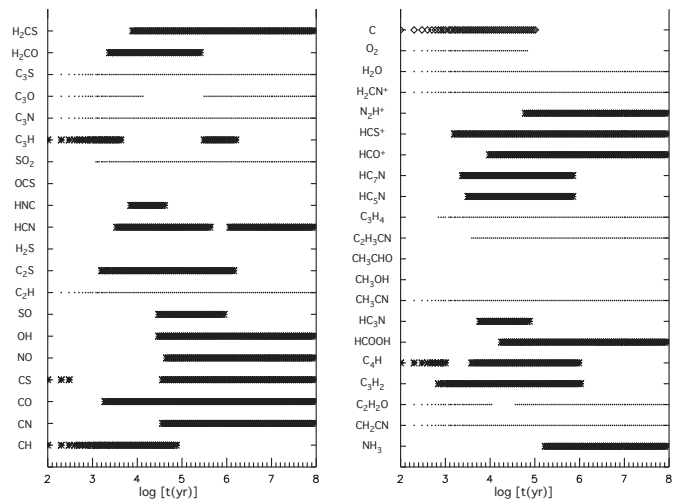}
\caption{Agreement between the observed abundances towards L134N ``N" peak) and the predictions of the chemical model (Model 2) as a function of time. Symbols indicate an agreement between the observational constraint and the model results. Stars are for the observed abundances whereas dots and diamonds are for the observational upper and lower limits respectively.  }
\label{Comp_L134N}
\end{center}
\end{figure}

%\begin{figure}
%\begin{center}
%\includegraphics[angle=90,width=1\linewidth]{age_cloud.eps}
%\caption{Percentage of molecules reproduced by Model 2 for L134N (solid line) and by Model 3 for TMC-1 (dashed line) as a function of time.}
%\label{age_cloud}
%\end{center}
%\end{figure}

\begin{table}
\caption{Fraction of species reproduced by the different models and databases with the mean time of best agreement (expressed logarithmically). \label{sum_fact}}
\begin{center}
\begin{tabular}{llccc}
\hline
\hline
  \multicolumn{5}{c}{Factor 3} \\
 \hline
 & & Model 1 & Model 2 & Model 3 \\
 \hline
 L134N & osu.2003 & 73\%, 5.7 & 80\%, 4.8  & 78\%, 4.7 \\
 & RATE99 & 78\%, 5.8  & 83\%, 5.9 & 66\%, 4.7 \\
 TMC-1 & osu.2003& 57\%, 4.4 & 61\%, 4.4 & 76\%, 4.9 \\
 & RATE99 & 68\%, 5.2  & 70\%, 5.7 & 70\%, 5.5 \\
 \hline
  \multicolumn{5}{c}{Factor 5} \\
 \hline
 & & Model 1 & Model 2 & Model 3 \\
 \hline
 L134N & osu.2003 & 82\%, 4.6  & 87\%, 4.7 & 85\%, 4.8 \\
 & RATE99 & 83\%, 4.8 & 83\%, 6.0 & 74\%, 5.2 \\
 TMC-1 & osu.2003& 61\%, 4.4 & 67\%, 4.4 & 86\%, 4.9 \\
 & RATE99 & 72\%, 5.6 & 78\%, 5.6 & 76\%, 5.3 \\
\hline
\end{tabular}
\end{center}
\end{table}%

Figure~\ref{Comp_L134N} shows the times of agreement for each observed molecule towards L134N. With our standard uncertainty of a factor of three in the observed abundances, Model 2 reproduces up to roughly 80\% of the 41 observed molecules in the range of times $(6.0-6.8)\times 10^4$~yr and the level of agreement tapers off gradually at both younger and older ages. The peak number increases to 87\% if an error of a factor of 5 is taken for the observed values. Model 1 reproduces fewer species (73\%), and the best agreement occurs later, at $5\times 10^5$~yr (see Table~\ref{sum_fact}).  With Model 2, the molecules H$_2$S, OCS, CH$_3$OH,  CH$_3$CHO are underestimated at all times. On the other hand, the observed abundances of HNC, C$_3$H, C$_3$O and NH$_3$ are reproduced by the model but not at the best range of ages.  The three species C$_{3}$H,  HNC, and NH$_{3}$ are in agreement with observation for times very close to the best range, while C$_{3}$O has not been detected and only an upper limit is known. The special cases of O$_{2}$ and H$_{2}$O, for which upper limits have been well studied,  are discussed in Sect.~\ref{O2_H2O}.

\subsection{Results for TMC-1}

\begin{figure}
\begin{center}
\includegraphics[width=1\linewidth]{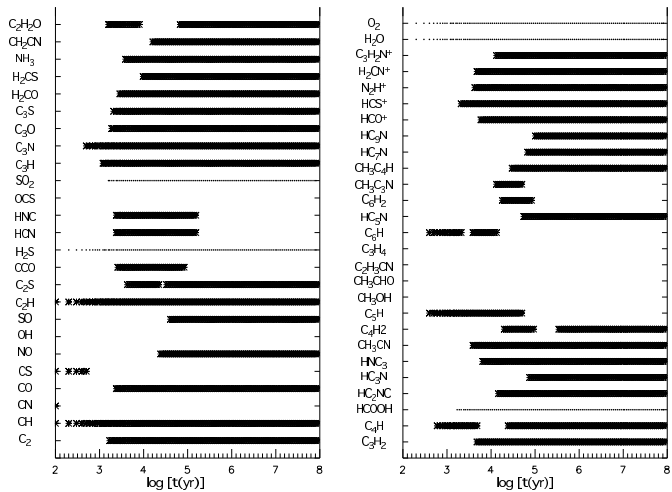}
\caption{Agreement between the observed abundances towards TMC-1 (``CP" peak) and the predictions of the chemical model (Model 3) as a function of time. }
\label{Comp_TMC1}
\end{center}
\end{figure}

Using Model 2, we can only reproduce at best 50\% of the observed 52 molecular abundances in the CP peak of TMC-1. This percentage is even lower than the 67\% obtained by \citet{2004MNRAS.350..323S}, who compared the results of the osu.2003 database with the same observations towards TMC-1CP. These authors considered  the modeling to be successful if the differences between the observed and theoretical values are less than a factor of 10.
 \citet{2004MNRAS.350..323S}, \citet{1998ApJ...501..207T} and \citet{2002A&A...395..233R} showed that by increasing the C/O elemental ratio, they were able to better reproduce the observations.  We then ran a third model (Model 3, see Table~\ref{diff_mod}) identical with Model 2 except that the initial abundance of O is lowered to  $1.2\times 10^{-4}$, so that the elemental C/O ratio becomes 1.2.  Fig.~\ref{Comp_TMC1} shows the comparison with Model 3. Here, seventy-six percent of the molecules are reproduced considering an observed uncertainty of a factor of 3 at a time of $8\times 10^4$~yr. This number is increased to 86\% at the same age for an observed uncertainty of a factor of 5. The molecules OH, OCS, CH$_3$OH, CH$_3$CHO, C$_2$H$_3$CN and C$_3$H$_4$ are never reproduced by the model whereas the CN, CS, C$_5$H, HC$_9$N, C$_6$H and CH$_3$C$_3$N abundances are not reproduced in the optimum age range. The observed abundance of OH in TMC-1 is quite uncertain since it was detected using a significantly larger beam than for the other species, and contamination from other regions of the cloud is probable \citep{1992IAUS..150..171O}.  

%\citet{2004MNRAS.350..323S} compared the results of the osu.2003 database introduced in %Sect.~\ref{model} with the same observations towards TMC-1CP. The authors considered  the %modeling to be successful if the differences between the observed and theoretical values were less %than a factor of 10. Contrary to \citet{2004MNRAS.350..323S} results, we found that the observed %abundances of C$_2$O and SO$_2$ (which are only upper limits) can be reproduced by Model 3.  %We have however an additional number of species for which the model fails to reproduce the %abundances:  CN, CS, OCS, C$_5$H, C$_3$H$_4$ and C$_6$H. 

Looking at both clouds, we see that the abundance of the molecules OCS, CH$_3$OH and CH$_3$CHO are underestimated at all times by the model compared with the observed values. One explanation could be these species are formed on the grains and released in the gas phase by non-thermal processes  \citep{2000ApJ...535..256M}. Indeed, OCS and CH$_3$OH are known to be present on the grain mantles with an abundance (comparable to that of H$_2$) of $\sim 10^{-7}$ for OCS \citep{1997ApJ...479..839P} and $\sim 10^{-6}$ for CH$_3$OH \citep{1996ApJ...472..665C}.  Work in progress by \citet{2006Garrod} with a gas-grain model and various non-thermal desorption mechanisms supports this view. The fact that H$_2$S and NH$_3$ are not reproduced correctly in L134N may have the same origin since H-enriched molecules are believed to form efficiently on grains although solid H$_2$S has never been detected on grains \citep{1998ARA&A..36..317V}.   For H-poor  species such as CS, CN and HNC, some of the rate coefficients of the critical reactions forming and depleting them may be more uncertain than we have assumed, especially if the reactions have not been studied and the estimated uncertainties are hiding non-random errors. 

 For the case of the two cold cores studied using osu.2003, as opposed to our previous study of hot cores, we find that with our sensitivity technique it is most often difficult to isolate small numbers of very important reactions for those molecules with abundances we cannot reproduce.  Rather, the general picture is that for many species, there are large numbers of reactions that are of relatively equal importance.   

 There are some reactions deemed important by our sensitivity technique that are critical for all the species: these are the ionization of H$_2$ and He by cosmic rays, as already noticed by \citet{2004AstL...30..566V}and \citet{Wakelam2005}, and raised in an oral presentation by A. Markwick-Kemper in the 16th UCL Astronomy Colloquium (Windsor, UK, 2002).  Ionization reactions involving cosmic rays, both direct and indirect, are, however, better thought of as variable parameters rather than reactions with uncertain rates since it not in general the physics of the process but the flux of cosmic rays that is in doubt.

\subsection{RATE99 calculations}

Before definitively ascribing the disagreements to specific causes, it is useful to see how the comparison with observations is affected by the use of the RATE99 network. Table~\ref{sum_fact} lists the overall level of agreement for both networks.  On balance, the results with RATE99 show slightly better agreement with observation and show it at later times.  Unlike the osu.2003 results, the RATE99 results are in agreement for the saturated species methanol and, in some cases, acetaldehyde.  Although this agreement would appear to weaken our argument that the saturated species are produced on grain surfaces, it should be noted that the methanol prediction of RATE99 is almost certainly wrong because the gas-phase synthesis of methanol used:
\begin{equation}
{\rm CH_{3}^{+} + H_{2}O \longrightarrow CH_{3}OH_{2}^{+} + h\nu, }
\end{equation}
\begin{equation}
{\rm CH_{3}OH_{2}^{+} + e^{-} \longrightarrow CH_{3}OH + H,}
\end{equation}
is assumed to be far too efficient for at least three reasons: (i) the radiative association reaction to produce the precursor ion CH$_{3}$OH$_{2}^{+}$ has been measured to occur much more slowly than predicted \citep{2002Lucas}, (ii) the dissociative recombination of this ion leads to methanol in only 6\% of collisions \citep{2006Geppert}, and (iii) the water abundance used in the radiative association reaction is far greater than its measured upper limit.  New calculations using the RATE99 network now agree with the osu.2003 result that this species cannot be produced in the gas \citep{2006Geppert}.  

The case of acetaldehyde may or may not be similar.  This species is also thought to be produced via a radiative association
\begin{equation}
{\rm H_{3}O^{+} + C_{2}H_{2} \longrightarrow CH_{3}CHOH^{+} + h\nu,}
\end{equation}
 followed by a dissociative recombination \citep{1987ApJ...313..867H}:
 \begin{equation}
 {\rm CH_{3}CHOH^{+} + e^{-} \longrightarrow CH_{3}CHO + H.}
 \end{equation}
There is no direct evidence for either reaction, although the association does occur at high densities in the laboratory via collisional stabilization \citep{1987ApJ...313..867H}.

We conclude that there is substantial evidence that methanol is only produced on grains, although the case against gas-phase production of acetaldehyde is not quite proven. It is likely though that the high abundance of acetaldehyde in the RATE99 calculation stems at least partially from an overly slow destruction by ions since this species has a dipole moment. The percentages of species in agreement using the RATE99 database listed in Table~\ref{sum_fact} do not include the agreement with these two species.

% {\bf Note that using the rate99 rate coefficients, the observations in both clouds are better reproduced %(seeTable~\ref{sum_fact}) at later times. The small variation of the physical parameters introduced in %Model 2 and 3 helps in reproducing the observed values. }

\section{Discussion}

\subsection{Special cases of water and molecular oxygen}\label{O2_H2O}

 \begin{figure}
\begin{center}
\includegraphics[width=1\linewidth]{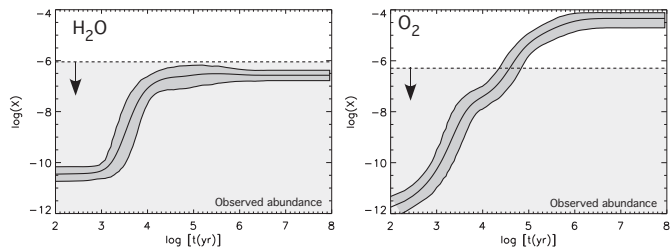}
\caption{Comparison between the theoretical abundances with error bars of H$_2$O and O$_2$ (Model 2) and the upper limits of the abundances found in L134N (see Table~\ref{ab_L134N}) }
\label{l134n_h2o_o2}
\end{center}
\end{figure}

An interesting result of this work is that it provides another manner in which the low upper limits to the abundances of gas-phase H$_2$O and O$_2$ in cold cores, detected by the SWAS and Odin satellites,  can be explained \citep{2000ApJ...539L.101S,2003A&A...402L..77P}.  Essentially, the upper limit of the observed abundance (multiplied by a factor of three given our standard chosen observational uncertainties) must not be less than the lower error bar of the theoretical abundance.  

Let us first consider the case of L134N.   In Fig.~\ref{l134n_h2o_o2}, we plot the calculated Model 2 abundances for water and oxygen with error bars as functions of time and superimpose the measured upper limits multiplied by a factor of three.  The results show that both species are successfully reproduced in the optimum age range defined by the maximum agreement with all species, although the case of O$_{2}$ is much more marginal since it is clearly overproduced at times after  $10^5$~yr.   This result shows that interaction with grain surfaces is not necessarily required if we consider young clouds ($\le 10^5$~yr).   The abundance of O$_2$ was found, however,  to be lower than the value for L134N  ($<10^{-7}$) in a dozen dark clouds \citep{2003A&A...402L..77P}, which we expect to have different ages. It may then be necessary to invoke depletion processes to explain the low abundance of O$_2$ in quiescent cores \citep[see][]{2002A&A...395..233R}.  The case of TMC-1 is singular since we used a model (Model 3) with high elemental C/O ratio.  Here there are no age constraints (see results in Fig.~\ref{Comp_TMC1}).

\subsection{Age of the clouds: a more refined estimate}\label{cloud_estimate}

 \begin{figure}
\begin{center}
\includegraphics[width=0.5\linewidth]{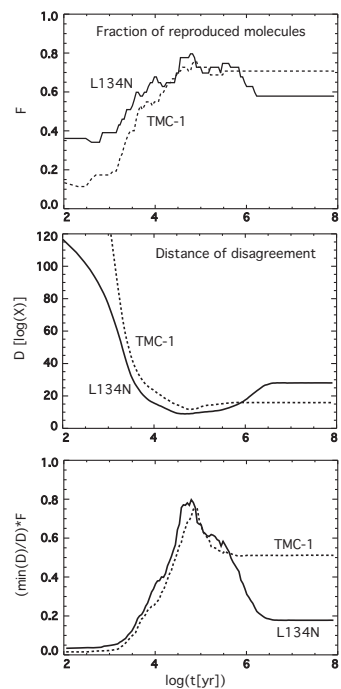}
\caption{Upper panel:  the fraction of molecules reproduced by the model  as a function of time for both clouds (Model 2 for L134N and Model 3 for TMC-1). Middle panel: the distance  D of disagreement between the observed and modeled abundances (see text for details).  Lower panel:   the fraction of reproduced molecules multiplied by min(D)/D (minimum of the distance of disagreement divided by the distance of disagreement). We consider a factor of 3 uncertainty in the observed abundances. }
\label{age_agree}
\end{center}
\end{figure}

Our models of quiescent cores are 
clearly not the complete picture since dynamical forces are producing and destroying cores 
while chemistry occurs \citep{2005MNRAS.356..654G}.  For TMC-1 and L134N,  the creation and chemistry have been occurring to some extent simultaneously.  So, our pseudo-time-dependent calculations represent a crude approximation, with 
frankly an uncertain zero-time condition.  Even within this crude approximation, we have used a less than perfect criterion to determine the ``chemical''  age  for TMC-1CP and L134N by simply maximizing the number of molecules with calculated abundances and error bars in agreement with observed values. Essentially, we have made no allowance for the quality of the agreement and the extent of the disagreement for individual species.  The results of this simple method are re-plotted in the top panel of Fig.~\ref{age_agree} for L134N (Model 2) and TMC-1CP (Model 3) as the fraction F of molecules in agreement vs time.  In this representation, we can see that the best age for L134N is somewhat more distinct that for TMC-1CP, but that the optimum ages lie in what previously was known as ``early time'', well before steady-state conditions set in.   

To do somewhat better, we first  compute an arbitrary parameter called the distance of disagreement D, which is defined,  for the species not reproduced by the model, as the sum over all  species of the distance $d_i$ (in units of $\log X$, see Fig.~\ref{Comp_L134N_OH}) between the observational value and the theoretical one.  For example,  if for a particular species, $i$,  $X_{obs} > X_{model}$ then the contribution to D is given by the expression
\begin{equation}
d_{i} =  \log X_{\rm obs;min} - \log X_{\rm model;max}, 
\end{equation}
where the maximum and minimum values refer  to the error bars. The middle panel of Fig.~\ref{age_agree} shows the total distance D as a function of time for the two clouds. This plot  indicates that the disagreement with the model reaches a minimum, labeled min(D), somewhat after $10^4$~yr.  Finally, the lower panel of Fig.~\ref{age_agree} shows a plot vs time of the ratio between the minimum distance, min(D), and D, multiplied by the fraction of reproduced molecules, F.  This quantity, which includes the strength of individual disagreements for specific species,  tends to sharpen the best age, especially for L134N.  In particular, we obtain a sharp  maximum for L134N around $6\times 10^4$~yr and a less marked maximum for TMC-1CP around 10$^{5}$ yr. These numbers are not strikingly different from a variety of other estimates, but must be taken with extreme caution. The ages derived analogously with the RATE99 set of reactions are somewhat larger, $\sim 10^6$~yr.

\section{Conclusions}

In this paper, we have reported the use of our Monte Carlo approach to uncertainties in calculated abundances to study the gas-phase chemistry of dark clouds, in particular the well-studied sources TMC-1CP and L134N.  We have utilized models in which the uncertainties in both rate coefficients and physical conditions are included; the latter can be thought of as either the actual observational uncertainties or physical heterogeneities in the sources.  With a specific criterion for agreement between observation and theory in which the errors in both techniques must overlap, we find that most but not species detected in the sources are reasonably well accounted for at so-called ``early time,'' a far from novel result although one determined more rigorously than in previous approaches.  

A major goal of the study has been to gain some insight as to which failing of the simple picture utilized is the more important: the lack of grain-surface chemistry or the lack of a dynamical model of the sources.  For saturated (H-rich) species, with reasonably understood syntheses on grain surfaces,  our results support earlier indications that surface chemistry is an important, probably a critical omission, but for unsaturated species, the picture remains less clear because it is difficult to determine from our sensitivity analysis whether the discrepancies can be accounted for by poorly determined rates of critical chemical reactions.  This difficulty stems from the  fact that the chemistry of unsaturated species often seems to involve contributions from large numbers of poorly determined reaction rates.  

The fact that discrepancies for the calculated abundances of individual species  between two chemical networks used, our osu.2003 network and the RATE99 network, can be larger than their calculated random uncertainties indicates that more attention must be paid in the laboratory to specific classes of reactions, such as ion-polar neutral systems,  before more definitive conclusions can be reached.  Further complicating the issue are other concerns such as the proper initial make-up of the gas, what elemental abundances are reasonable, and the range over which the cosmic ray ionization rate $\zeta$ can be varied.  Nevertheless, our study indicates strongly that gas-phase models of static sources cannot represent the complete picture.  More complex treatments, involving cycles of surface adsorption, reaction, and desorption, possibly coupled with dynamical histories of the sources, would appear to be required.

\begin{acknowledgements}

We thank the National Science Foundation for its support of the astrochemistry program at Ohio State.

\end{acknowledgements}

\bibliographystyle{aa}

%\bibliography{aamnem99,biblio}

\end{document}